\documentclass[10pt,twocolumn,letterpaper]{article}

\usepackage{iccv}
\usepackage{times}
\usepackage{epsfig}
\usepackage{graphicx}
\usepackage{amsmath}
\usepackage{amssymb}
\usepackage[accsupp]{axessibility}
\usepackage{makecell}
\usepackage{multirow}
\usepackage{adjustbox}
\usepackage{hhline}  
\usepackage{booktabs}
\usepackage{tablefootnote}
\usepackage{threeparttable}
\usepackage{bm}
\usepackage{hyperref}
\usepackage{algorithm}
\usepackage{algorithmic}
\hypersetup{
    colorlinks=true,
    linkcolor=blue,
    filecolor=magenta,      
    urlcolor=cyan,
    pdftitle={Overleaf Example},
    pdfpagemode=FullScreen,
    }

\usepackage{color}

\iccvfinalcopy 

\def\iccvPaperID{3124} 

\ificcvfinal\fi

\begin{document}

\title{RestoreDet: Degradation Equivariant Representation for Object Detection in Low Resolution Images}





\author{Ziteng Cui\textsuperscript{1},
    Yingying Zhu\textsuperscript{2},
    Lin Gu\textsuperscript{3,4}\thanks{Corresponding author.}, 
    Guo-Jun Qi\textsuperscript{5},
    Xiaoxiao Li\textsuperscript{6},\\
    Peng Gao\textsuperscript{7},
    Zenghui Zhang\textsuperscript{1},
    Tatsuya Harada\textsuperscript{4,3}\\
    \textsuperscript{1}Shanghai Jiao Tong University, \textsuperscript{2}University of Texas at Arlington, \textsuperscript{3}RIKEN AIP, \textsuperscript {4} The University of Tokyo\\ \textsuperscript{5}Innopeak Technology, 
    \textsuperscript {6} The University of British Columbia \textsuperscript {7} Shanghai AI Laboratory
    }

\maketitle
\ificcvfinal\thispagestyle{empty}\fi

\begin{abstract}
Image restoration algorithms such as super resolution (SR) are indispensable pre-processing modules for object detection in degraded images. However, most of these algorithms assume the degradation is fixed and known a priori. When the real degradation is unknown or differs from assumption, both the pre-processing module and the consequent high-level task such as object detection would fail. Here, we propose a novel framework, RestoreDet, to detect objects in degraded low resolution images. RestoreDet utilizes the downsampling degradation as a kind of transformation for self-supervised signals to explore the equivariant representation against various resolutions and other degradation conditions. Specifically, we learn this intrinsic visual structure by encoding and decoding the degradation transformation from a pair of original and randomly degraded images. The framework could further take the advantage of advanced SR architectures with an arbitrary resolution restoring decoder to reconstruct the original correspondence from the degraded input image. Both the representation learning and object detection are optimized jointly in an end-to-end training fashion. RestoreDet is a generic framework that could be implemented on any mainstream object detection architectures. The extensive experiment shows that our framework based on CenterNet has achieved superior performance compared with existing methods when facing variant degradation situations. Our code would be released soon.
\end{abstract}


\section{Introduction}
High level vision tasks (\textit{i.e.} image classification, object detection, and semantic segmentation) have witnessed great success thanks to the large scale dataset \cite{imagenet, coco_dataset, voc_dataset}. Images in these datasets are  mainly captured by commercial cameras with higher resolution and signal-to-noise ratio (SNR). Trained and optimized on these high-quality images, high-level vision would suffer a performance drop on low resolution ~\cite{SR_for_vision_tasks, SR_object_detection, SR_for_segmentation} or low quality images~\cite{noise_4_vision,Blur_4_vision, CVPR_blur_detection,White_balance,CVPR_low_face, ICCV_MAET}.

To improve the performance of vision algorithms on degraded low resolution images,  Dai \textit{et al.}~\cite{SR_for_vision_tasks} presented the first  comprehensive study advocating pre-processing  images with super resolution (SR) algorithms. Other high-level tasks like face recognition~\cite{low_face_recognition}, face detection~\cite{Low_face_CVPR19}, image classification~\cite{low_image_classification, noise_4_vision} and semantic segmentation~\cite{SR_for_segmentation}, also benefit from the restoration module to extract more discriminate features.

Most existing enhancement methods, especially SR algorithms~\cite{SR_kernelGAN,SR_USRNet,SR_cvpr2021_random}, assume target  images are from a \textbf{known and fixed} degradation model~\cite{Degradation_model,Degradation_model_1}: 
\begin{equation}
    t(x) = (x \circledast k)\downarrow_s + n,
    \label{eq:degradation_model}
\end{equation}
where $t(x)$ and $x$  denote the degraded low resolution (LR) image and original high resolution (HR) input respectively. $k$ is the blur kernel while $\downarrow_s$ is the down-sampling operation with ratio $s$.  $n$ is the additive noise. However, the performance of these enhancement algorithms would decline severely when the real degradation deviates from the assumption~\cite{SR_Iter_Kernel}. To make it worse, since these restoration methods are human visual perception oriented, they are often inadequate for machine perception tasks such as object detection~\cite{ICCV_MAET,TIP20_lowvisibility,resizer_2021_ICCV}.

\begin{figure}
    \centering
    \includegraphics[width = 8.7cm, height = 6cm]{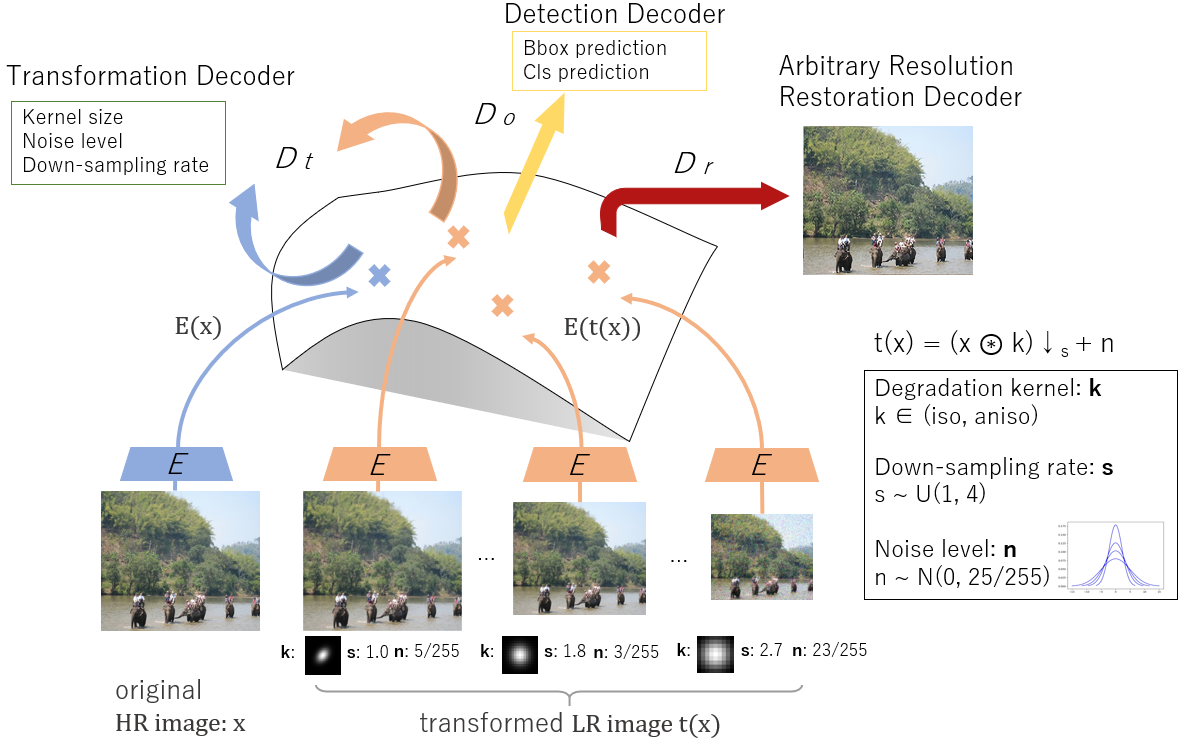}
    \caption{A simple illustration of our framework. Encoder $E$ encodes the image into degradation equivariant representations. Three decoders $D_t$, $D_o$ and $D_r$ then decode the representations for transformation prediction, object detection and image restoration respectively.}
    \label{fig:repre_manifold}
\end{figure}

Instead of explicitly enhancing an input image with a restoration module under strict assumptions, we exploit the intrinsic equivariant representation against various resolutions and degradation status. Based on the encoded representation shown in Fig.\ref{fig:repre_manifold}, we propose RestoreDet, an end-to-end model for object detection in degraded LR images. To capture the complex patterns of visual structures, we utilize groups of downsampling degradation transformations as the self-supervised signal~\cite{ICCV_MAET,aet}. During the training, we generate a degraded LR image $t(x)$ from the original HR image $x$ through a random degradation transformation $t$. As shown in Fig.\ref{fig:repre_manifold}, this pair of images are fed into the encoder $E$ for their latent feature  $E(x)$ and $E(t(x))$. To train the Encoder $E$ to learn the degradation equivariant representation, we at first introduce a transformation decoder $D_t$ to decode the applied degraded transformation $t$ from the representation $E(x)$ and $E(t(x))$. If the transformation could be reconstructed, the representation should capture the dynamics of how they change under different transformations as much as possible~\cite{group_convolution,capsule_network,aet}.

To further take advantage of the strength of the fast-growing SR research, we introduce an arbitrary-resolution restoration decoder (ARRD) $D_r$ Fig.\ref{fig:repre_manifold}. ARRD reconstructs the original HR data $x$ from the representation $E(t(x))$ of various degradated LR image $t(x)$. ARRD $D_r$ would supervise the encoder $E$ to encode the detailed image structure which facilitates the consequent tasks. Based on the encoded representation $E(t(x))$, the object detection decoder $D_o$ then performs the detection to get the object's location and class. During inference, the target image is directly passed through the encoder $E$ and object detection decoder $D_o$ in Fig.\ref{fig:repre_manifold} for detection. Compared to pre-processing module based methods~\cite{Aerial_detection_SR, SR_object_detection},  our inference pipeline is more computation efficient as we avoid explicitly reconstructing the image details.


 To cover the diverse degradation in real scenario, we generate degraded $t(x)$ by randomly  sampling a transformations $t$ according to practical down-sampling degradation  model~\cite{Degradation_model_1, SR_cvpr2021_random}. As shown in Fig.\ref{fig:repre_manifold},  the transformation $t$ is characterized by down-sampling ratio $s$, degradation kernel $k$,  and noise level $n$ in Eq.\ref{eq:degradation_model}.

Our contribution could be summarized as follows:

\begin{itemize}
    \item We propose a novel framework, RestoreDet, to detect objects in degraded low resolution images. To achieve this, we learn the degradation equivariant representation that captures the dynamics of feature representations under  diverse resolutions and degradation types. RestoreDet is a generic framework that could be integrated into standard object detection architecture. 
    
   \item Our method takes the strength of advanced super resolution (SR) research by training an arbitrary resolution restoration decoder (ARRD) that reconstructs the high resolution details. Furthermore, by optimizing the representation learning and detection in a unified end-to-end training framework, the representation preserves the intrinsic visual structure that is discriminative for detection.

    \item We evaluate our method on two mainstream public datasets KITTI~\cite{kitti_dataset} and MS COCO~\cite{coco_dataset}. The experiment results also show that our method has achieved SOTA performance on all low resolution object detection tasks (down $1\sim4$, down $2$, down $4$). \textbf{We will release the source code and results on Github  upon the acceptance of this submission.}
\end{itemize}

\section{Related Works}

\subsection{Single Image Super Resolution}

The very first CCN-based SR was proposed in Dong \textit{et al.} ~\cite{SRCNN} with a three-layer neural network. Then Kim \textit{et al.} ~\cite{SR_deepCNN} extended the depth of network to 20 layers with gradient clipping  and residual learning. Batch normalization is later identified to impose the negative effect  on the HR reconstruction. By removing this layer, Lim \textit{et al.} proposed enhanced deep super resolution network (EDSR) ~\cite{SR_EDSR}  that achieves the SOTA in 2017. After ESDR, better SR architectures are designed by  integrating the successful deep learning techniques such as Laplacian pyramid structure ~\cite{SR_Laplacian}, dense connection~\cite{SR_densenet},  residual dense connection ~\cite{SR_RDN_block}, graph neural network ~\cite{SR_Graph_NN} and so on. Besides designing sophisticated architecture, losses like perceptual loss ~\cite{SR_perpectual_loss},  adversarial loss ~\cite{SRGan} and targeted perceptual loss ~\cite{SR_target_perpectual_loss} are also demonstrated to improve the SR reconstruction quality. 

SR algorithms heavily rely on the assumption of degradation model. Much efforts are spent to relax the constraint. Not limited to a fixed integral upsampling scale, Hu \textit{et al.} ~\cite{SR_arbitary_MetaSR} proposed a meta-upscaling factor to super-resolve the image with  arbitrary scale factor.  Recently, Fu \textit{et al.} ~\cite{Fu_Neurocomp2021} proposed a solution based on residual attention network.

To deploy SR for real scenarios, blind SISR methods assume the degradation information is not known. One direction is to convert the problem into non-blind SR by initially estimate the degradation parameters~\cite{SR_kernelGAN}. However, the applied non-blind SR algorithm~\cite{ZSSR_CVPR18} is very sensitive to the error of the degradation estimation.  Gu \textit{et al.}  ~\cite{SR_Iter_Kernel} then proposed to iteratively correct the estimated degradation with an iterative kernel correction (IKC) method. Without explicitly estimating degradation parameters,  Wang \textit{et al.} ~\cite{Wang_2021_CVPR_contrastiveSR} introduced a contrastive loss to design the Degradation-Aware SR network based on the learned representations. Recently, Zhang \textit{et al.} ~\cite{SR_cvpr2021_random} solved the general blind SISR by designing a practical model considering complex degradations. This model has been demonstrated to cover the degradation space of real images. Therefore, we also adopt this practical model to synthesize various degraded LR images as the self-supervised signal to train our model.



\subsection{Image Restoration for Machine Perception}

There is sufficient evidence that pre-processing module is an effective in improving the performance of high-level tasks on degraded images~\cite{TIP20_lowvisibility,resizer_2021_ICCV}. Dai \textit{et al.} ~\cite{SR_for_vision_tasks} made the first extensive analysis on improving several vision tasks with SR as pre-process. Wang \textit{et al.} ~\cite{low_image_classification} analyzed the effectiveness of SR in image classification task while DSRL ~\cite{SR_for_segmentation} improved the semantic segmentation with an additional SR block. Shermeyer and Etten~\cite{Aerial_detection_SR}  evaluate the  effectiveness of a SR pre-process step on aerial image object detection. Recently, Haris \textit{et al.} ~\cite{SR_object_detection} jointly optimized the SSD detection loss~\cite{detection_SSD} along with SR sub-network~\cite{SR_DBPN} to improve detection performance.

Similarly, the denoising algorithms have also be explored in  ~\cite{noise_4_vision,robust_segmentation_1,CVPR_blur_detection}.   Kamann \textit{et al.}~\cite{robust_segmentation_1} studied the impact of noise and blur on different semantic segmentation methods.  Liu \textit{et al.}~\cite{noise_4_vision} combined a denoise network in classifier to improve classification's performance.  Very recently, Mohamed and Gabriel~\cite{CVPR_blur_detection}  analyzed  motion blur to  improve detection performance on motion blurry images.

 However, most of these existing works assume the degradation parameters such as the downsampling ratio is known. Based on the degradation equivariant representation, our proposed framework is robust to various unknown degradation in real-world scenarios. Without an explicit restoration module, we directly perform the detection on low-dimension encoded features that saves much computational burden.

\section{Downsampling Degradation Transformations}
\label{sec:degradation}


\begin{figure*}
    \centering
    \includegraphics[width = 16cm, height = 8cm]{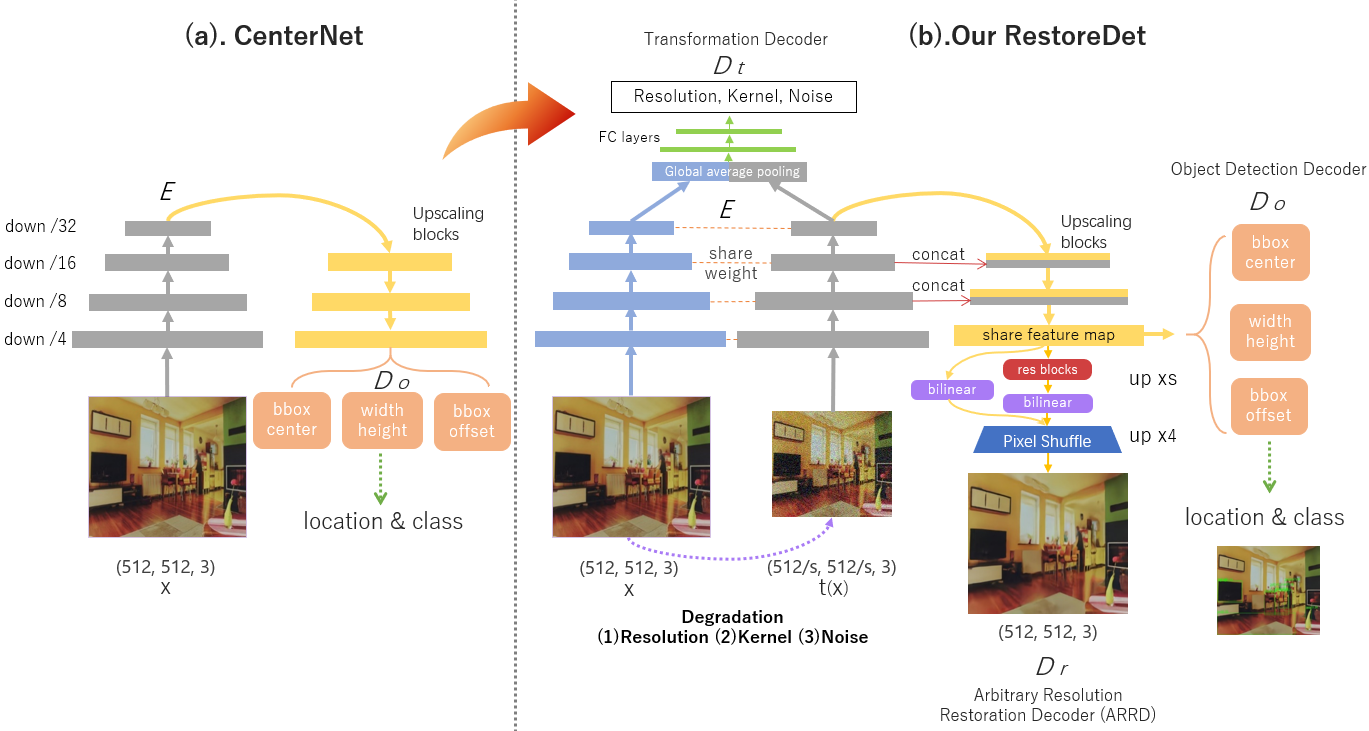}
    \caption{An illustration of how our RestoreDet is implemented based on CenterNet.  Left is the original CenterNet~\cite{centernet} while the right one is the architecture of our  RestoreDet.}
    \label{fig:detail_structure}
\end{figure*}

In real scenarios, the image may be captured and processed in various ways. To cover these generally unknown operations, it is necessary to select a practical degradation model for the degradation transformation. This model would  transform the high resolution (HR) image $x$ to the low resolution  counterpart $t(x)$ with Eq.~\ref{eq:degradation_model}.

Early enhancement methods assume a simple degradation model where LR is directly downsampled from the HR images without or with simple noise. Instead of dealing with synthetic images, more and more methods now focus on more realistic degradation models. For example, \cite{SR_real_images_1, SR_real_images_2} directly train their model on the  LR and HR images pair captured by the real camera system.  USRNet~\cite{SR_USRNet} effectively handled the degradation models with different parameters such as scale factors by unfolding the model-based energy function. Therefore, we adopt the practical degradation model~\cite{Degradation_model,Degradation_model_1,SR_cvpr2021_random} that accounts for diverse degradation in real images.





Our degradation process in Eq.\ref{eq:degradation_model} mainly comprises of three parts: blur kernel $k$, down-sampling ratio $s$ and additional noise $n$. During training, $k$, $s$, $n$  is randomly sampled from the corresponding distributions as follows:



\textbf{Convolution Operation:} Gaussian blur kernel is the most common  kernel to blur the image~\cite{gaussian_kernel, gaussian_kernel_2, SR_kernelGAN, SR_UDVD}. Here, we choose two Gaussian degradation kernels: isotropic Gaussian kernels $k_{iso}$ and anisotropic Gaussian kernels $k_{aniso}$~\cite{SR_kernelGAN, SR_SRMD,  SR_cvpr2021_random}. We also consider none degradation kernel $k_{none}$. Following ~\cite{SR_cvpr2021_random}, the kernel size is uniformly sampled from $\left\{7\times7, 9\times9, ..., 21\times21 \right\}$ and the $k_{iso}$'s width is uniformly chosen from $[0.1, 2.4]$. For $k_{aniso}$, the kernel angle is uniformly chosen from $[0, \pi]$ and the longer kernel width is uniformly chosen from $[0.5, 6]$. Here, the transformation decoder $D_{t}$ is trained to learn the kernel size. During training, we sample different kernel size, width, and types with the equal probability.

\textbf{Down-sampling:} For the down-sampling process, the sampling ratio $s$ is randomly chosen from uniform distribution $U(1, 4)$ (\textit{e.g.} 2.2) while the down-sampling  methods is randomly chosen from nearest method $d_{nearest}^s$, bilinear method $d_{bilinear}^s$ and bicubic method $d_{bicubic}^s$. In the actual training phase, the down-sampled resolution must be multiple of the network's max downsampling rate (\textit{e.g.} 32 for ResNet~\cite{resnet}\footnote{A $512 \times 512$ image $x$. For example, the degraded image $t(x)$'s resolution must be integral multiple of $32$, so the down-sampling rate $s$ is chosen from $[\frac{512}{128}, \frac{512}{160}, \frac{512}{192}, ..., \frac{512}{512}$].}), and the resolution in each batch should be same. Here the transformation decoder $D_{t}$ is responsible for decoding the scale factor $s$ to make the resolution prediction. 

\textbf{Noise:} When dealing with the real-world scenarios,  the Gaussian noise model is usually adopted to simulate the noises from  camera sensor noise ~\cite{camera_sensor_noise}, low-light noise ~\cite{low_light_noise} to quantization noise ~\cite{quantisation_noise}), \textit{etc.} Therefore,  we adopt a zero-mean additive white Gaussian noise (AWGN) model as $n$ in Eq.\ref{eq:degradation_model}. The variance $\sigma$ is randomly chosen from a uniform distribution $U(0, 25/255)$ (\textit{e.g.} $13.3/255$).



\section{RestoreDet}

Our RestoreDet is built upon CenterNet~\cite{centernet} due to its concise structure. We emphasize that RestoreDet is a generic framework that could also be implemented on other popular object detectors~\cite{detection_yolo,detection_yolof,DETR,SMCA_DETR} and  backbones~\cite{mobile_net,gao2021container}. 

\subsection{CenterNet}

CenterNet~\cite{centernet} is an efficient one-stage anchor-free object detector. We show its vanilla  structure of~\cite{centernet} in Fig.\ref{fig:detail_structure} (a). Input image is fed to the backbone, \textit{i.e.}  ResNet18~\cite{resnet}, to extract $/32$ bottleneck feature, and then  upsampled to a $/4$ feature map by three $\times2$ deconvolution blocks.  This $/4$ feature map is  passed to scaling blocks with three independent convolution blocks to generate  the final feature maps. Based on this feature map,  there are prediction heads conducting class-wise bbox center detection, bbox height and width regression, offset regression respectively. For more details, please refer to ~\cite{centernet}.


The CenterNet could be decomposed into an encoder-decoder style structure. Here, we denote the network backbone part (gray part in Fig.\ref{fig:detail_structure}.a) as encoder $E$. The object detection decoder $D_o$, comprised by three prediction heads (colored in orange in Fig.\ref{fig:detail_structure} (a)), decodes the object information.




\subsection{Architecture and Training Pipeline}

 Fig.\ref{fig:detail_structure} (b) illustrates how to implement our RestoreDet based on CenterNet~\cite{centernet}. The detailed training procedure is given in Algo.\ref{RestoreDet_Algorithm}. When training the RestoreDet, the original HR image $x$ and transformed degraded LR image $t(x)$ are sent to the encoder $E$ to encode the degradation equivariant representation. Here, we directly use the encoder $E$ of CenterNet but duplicate it into a shared weight Siamese structure to receive the  HR and LR image, respectively.

 With this encoder $E$, the encoded representative $E(x)$ and $E(t(x))$ would have same channel but different size. Then we apply a global average pooling layer $G$ to pool them into the same shape 2-D tensor.  $G(E(x))$ and $G(E(t(x)))$ are concatenated together for the transformation decoder $D_t$, which comprises only two FC layers. $D_t$ predicts the transformation parameters $\hat{k}$, $\hat{s}$, $\hat{n}$, as illustrated in Sec.~\ref{sec:degradation}.  $k$ represents the Gaussian kernel size, $s$ represents the down-sample factor and $n$ represents noise level. While training, all the ground truth $k$, $n$, $s$ is normalized to $0 \sim 1$ in their own type. For transformation decoder $D_t$, we have  MSE loss $l_{trans}$ between $(\hat{k}, \hat{s}, \hat{n})$ and $(k, s, n)$.
 

We only put representation $E(t(x))$ of LR image $t(x)$ into upscaling blocks with three $\times2$ deconvolution blocks to generate the final feature map for object detection decoder $D_o$ and arbitrary resolution restoration decoder (ARRD) $D_r$. As illustrated by the red arrows in Fig.\ref{fig:detail_structure}, we introduce the skip connection on  $/8$ and $/16$ feature maps between the backbone encoder $E$ and the deconvolution blocks. Fusing  features  from different scales could enhance semantic information and contribute to the subsequent $D_o$ and $D_r$.



We further regularize the representation encode $E$ with our unique ARRD $D_r$ that recovers HR image $\hat{x}$. Since the downsampling rate $s$ is not a fixed integral number in the training stage, ARRD could deal with an arbitrary scale factor. Transformation decoder $D_t$ and ARRD $D_r$ could force the encoder $E$ to not only  capture the dynamics of how images  change under different transformations, but also extracts the  complex patterns of visual structures. Since ARRD $D_r$ aims to recover  the original resolution of clean image $x$ from $E(t(x)))$, it could also  support the object detection decoder $D_o$ with more detailed features. Motivated by the learnable resizer model~\cite{resizer_2021_ICCV}, we design this decoder with a residual bilinear model shown in Fig.~\ref{fig:detail_structure}, which ends up with a $\times 4$ pixel shuffle layer~\cite{SR_pixel_shuffle}. ARRD is a light weight structure that uses fewer parameters (0.06M) compared to the backbone encoder (11.17M), upscaling blocks (3.61M) and detection decoder $D_o$ (0.12M). 
The ARRD loss $l_{d}$ is defined as an L1 loss between output image $\hat{x}$ and ground truth image $x$.

\begin{equation}
\begin{aligned}
    l_{d} &= |\hat{x} - x|_1 = |D_r(E(t(x))) - x|_1\\
    l_{trans} &= ||D_{t}\left[ E(t(x)), E(x)\right] - (k, s, n)||_2^2
\end{aligned}
\end{equation}

We adopt the three CenterNet prediction heads as the object detection decoder $D_o$ to conduct detection on the final feature map generated by the upscaling block.



As shown in Algo.\ref{RestoreDet_Algorithm}, we optimise the total loss $l_{total}$ including detection loss  (classwise bbox center loss, bbox width and height loss, bbox offset loss) $l_{obj}$,   transformation loss  $l_{trans}$ and data restoration loss $l_{d}$.
\begin{equation}
    l_{total} = l_{obj} + \lambda_1 \cdot l_{trans} + \lambda_2 \cdot l_{d},
\end{equation}
where $\lambda_1$ and $\lambda_2$ is the non-negative parameters for loss balancing. They  are respectively set to 8 and 0.8 in our experiments.

\subsection{Inference Procedure}

The inference procedure only involves  encoder $E$ , upscaling block and  object detection decoder $D_o$ as illustrated in Fig.\ref{fig:detail_structure}. Specifically, the encoder $E$ encodes the input target image before $D_o$ performs the detection. Compared to explicitly pre-processing image for high-level tasks~\cite{SR_for_vision_tasks, Aerial_detection_SR, SR_object_detection}, our RestoreDet saves much computational time as we avoid reconstructing HR details of data.



We could also reconstruct the HR image with our ARRD decoder $D_r$.  Very interestingly, our restored images $\hat{x}$ are more machine vision oriented and exhibit artifacts around the center of the object, as shown in Fig.\ref{fig:results}.

\begin{algorithm}[t]
\caption{RestoreDet Algorithm Pipeline}
\label{RestoreDet_Algorithm}
\textbf{(1). Data Generation}: \\
\small B: bacth size, C: channel, H: image height, W: image width\\
\normalsize \textbf{inputs:} HR image $x=(B, C, H, W)$, down-sample factor $s \sim (1.0, 4.0)$ \\
\textbf{outputs:} down-sampling degraded LR image $t(x)=(B, C, \frac{H}{s}, \frac{W}{s})$ 
\begin{algorithmic}
\FOR{$i$ in range($B$):}
\STATE (1). Convolution with blur kernel $k$
\STATE (2). Down-sampling with rate $s$
\STATE (3). Add noise $n$
\ENDFOR\\
\end{algorithmic}
\textbf{(2). Training}:\\
\textbf{inputs:} HR image $x=(B, C, H, W)$, and degraded LR image $t(x)=(B, C, \frac{H}{s}, \frac{W}{s})$ \\
\textbf{outputs:} detection output, estimated SR image $\hat{x}$, estimated transformation $\hat{t}$\\
\textbf{encoding:} \\
$ x \underrightarrow{\quad E \quad} E(x), t(x) \underrightarrow{\quad E \quad} E(t(x)) \\$
\textbf{decoding:} \\
transformation decoding: $\hat{t} = D_t([E(x), E(t(x))])$ \\
data restoration decoding: $\hat{x} = D_r(E(t(x))$ \\
detection decoding: detection results $= D_o(E(t(x))$
\end{algorithm}

\begin{table*}[]
\centering
\begin{adjustbox}{max width= 1 \linewidth}
\begin{tabular}{cccccccccc}
\toprule[1.2pt]
\hline
\multicolumn{1}{c|}{\multirow{2}{*}{model}}     & \multicolumn{1}{c|}{\multirow{2}{*}{train set}} & \multicolumn{2}{c|}{Pre-process to test set}                                             & \multicolumn{1}{c|}{KITTI (car)} & \multicolumn{4}{c|}{COCO (80 classes)}                                                                          & \multirow{2}{*}{FPS (COCO)} \\ \cline{3-9}
\multicolumn{1}{c|}{}                           & \multicolumn{1}{c|}{}                           & \multicolumn{1}{c|}{method}                   & \multicolumn{1}{c|}{up-resolution}       & \multicolumn{1}{c|}{$\rm{AP_{50}}$}        & \multicolumn{1}{c|}{$\rm{AP}$}   & \multicolumn{1}{c|}{$\rm{AP_s}$}  & \multicolumn{1}{c|}{$\rm{AP_m}$}   & \multicolumn{1}{c|}{$\rm{AP_l}$}   &                             \\ \hline
\multicolumn{10}{c}{Detection Performance on Original Set (Upper Bound)}                                                                                                                                                                                                                                                                                                        \\ \hline
\multicolumn{1}{c|}{CenterNet}                  & \multicolumn{1}{c|}{N}                          & \multicolumn{1}{c|}{-}                        & \multicolumn{1}{c|}{-}                   & \multicolumn{1}{c|}{85.6}        & \multicolumn{1}{l|}{30.0} & \multicolumn{1}{l|}{10.6} & \multicolumn{1}{l|}{0.332} & \multicolumn{1}{l|}{0.472} & 51.0                        \\ \hline
\multicolumn{10}{c}{Interpolation Methods}                                                                                                                                                                                                                                                                                                                                      \\ \hline
\multicolumn{1}{c|}{\multirow{2}{*}{CenterNet}} & \multicolumn{1}{c|}{\multirow{2}{*}{N}}         & \multicolumn{1}{c|}{\multirow{2}{*}{bicubic}} & \multicolumn{1}{c|}{x2}                  & \multicolumn{1}{c|}{50.6}        & \multicolumn{1}{c|}{16.2} & \multicolumn{1}{c|}{4.1}  & \multicolumn{1}{c|}{15.3}  & \multicolumn{1}{c|}{31.1}  & 50.0                        \\ \cline{4-10} 
\multicolumn{1}{c|}{}                           & \multicolumn{1}{c|}{}                           & \multicolumn{1}{c|}{}                         & \multicolumn{1}{c|}{x4}                  & \multicolumn{1}{c|}{36.5}        & \multicolumn{1}{c|}{8.0}  & \multicolumn{1}{c|}{4.6}  & \multicolumn{1}{c|}{10.5}  & \multicolumn{1}{c|}{10.1}  & 16.2                        \\ \hline
\multicolumn{10}{c}{Normal Deep Super Resolution Methods}                                                                                                                                                                                                                                                                                                                       \\ \hline
\multicolumn{1}{c|}{\multirow{5}{*}{CenterNet}} & \multicolumn{1}{c|}{\multirow{5}{*}{N}}         & \multicolumn{1}{c|}{RCAN~\cite{SR_RCAN}}                     & \multicolumn{1}{c|}{\multirow{5}{*}{x2}} & \multicolumn{1}{c|}{47.5}        & \multicolumn{1}{c|}{13.8} & \multicolumn{1}{c|}{3.0}  & \multicolumn{1}{c|}{13.9}  & \multicolumn{1}{c|}{26.4}  & 50.2                        \\ \cline{3-3} \cline{5-10} 
\multicolumn{1}{c|}{}                           & \multicolumn{1}{c|}{}                           & \multicolumn{1}{c|}{HAN~\cite{SR_HAN}}                      & \multicolumn{1}{c|}{}                    & \multicolumn{1}{c|}{46.9}        & \multicolumn{1}{c|}{14.8} & \multicolumn{1}{c|}{3.0}  & \multicolumn{1}{c|}{14.8}  & \multicolumn{1}{c|}{27.9}  & 49.7                        \\ \cline{3-3} \cline{5-10} 
\multicolumn{1}{c|}{}                           & \multicolumn{1}{c|}{}                           & \multicolumn{1}{c|}{IMDN~\cite{Light_weight_SR_LIMN}}                     & \multicolumn{1}{c|}{}                    & \multicolumn{1}{c|}{51.3}        & \multicolumn{1}{c|}{15.0} & \multicolumn{1}{c|}{3.5}  & \multicolumn{1}{c|}{14.3}  & \multicolumn{1}{c|}{27.4}  & 49.6                        \\ \cline{3-3} \cline{5-10} 
\multicolumn{1}{c|}{}                           & \multicolumn{1}{c|}{}                           & \multicolumn{1}{c|}{PAN~\cite{SR_PAN}}                      & \multicolumn{1}{c|}{}                    & \multicolumn{1}{c|}{50.5}        & \multicolumn{1}{c|}{14.8} & \multicolumn{1}{c|}{3.3}  & \multicolumn{1}{c|}{14.1}  & \multicolumn{1}{c|}{27.8}  & 50.1                        \\ \cline{3-3} \cline{5-10} 
\multicolumn{1}{c|}{}                           & \multicolumn{1}{c|}{}                           & \multicolumn{1}{c|}{SWIN-IR~\cite{SR_SWIN-IR}}                  & \multicolumn{1}{c|}{}                    & \multicolumn{1}{c|}{51.6}        & \multicolumn{1}{c|}{15.2} & \multicolumn{1}{c|}{3.5}  & \multicolumn{1}{c|}{14.9}  & \multicolumn{1}{c|}{28.4}  & 50.2                        \\ \hline
\multicolumn{10}{c}{Real World Super Resolution Methods}                                                                                                                                                                                                                                                                                                                        \\ \hline
\multicolumn{1}{c|}{\multirow{4}{*}{CenterNet}} & \multicolumn{1}{c|}{\multirow{4}{*}{N}}         & \multicolumn{1}{c|}{Real-SR~\cite{SR_real_images_1}}                  & \multicolumn{1}{c|}{\multirow{4}{*}{x2}} & \multicolumn{1}{c|}{55.6}        & \multicolumn{1}{c|}{14.2} & \multicolumn{1}{c|}{3.1}  & \multicolumn{1}{c|}{12.4}  & \multicolumn{1}{c|}{29.5}  & 51.1                        \\ \cline{3-3} \cline{5-10} 
\multicolumn{1}{c|}{}                           & \multicolumn{1}{c|}{}                           & \multicolumn{1}{c|}{RRDB~\cite{RRDB}}                     & \multicolumn{1}{c|}{}                    & \multicolumn{1}{c|}{43.1}        & \multicolumn{1}{c|}{10.4} & \multicolumn{1}{c|}{2.8}  & \multicolumn{1}{c|}{11.9}  & \multicolumn{1}{c|}{20.3}  & 50.6                        \\ \cline{3-3} \cline{5-10} 
\multicolumn{1}{c|}{}                           & \multicolumn{1}{c|}{}                           & \multicolumn{1}{c|}{Blind-SR~\cite{Siggraph_blind_SR}}                 & \multicolumn{1}{c|}{}                    & \multicolumn{1}{c|}{64.2}        & \multicolumn{1}{c|}{15.9} & \multicolumn{1}{c|}{3.0}  & \multicolumn{1}{c|}{14.6}  & \multicolumn{1}{c|}{34.2}  & 49.8                        \\ \cline{3-3} \cline{5-10} 
\multicolumn{1}{c|}{}                           & \multicolumn{1}{c|}{}                           & \multicolumn{1}{c|}{BSRGAN~\cite{SR_cvpr2021_random}}                   & \multicolumn{1}{c|}{}                    & \multicolumn{1}{c|}{70.8}        & \multicolumn{1}{c|}{16.8} & \multicolumn{1}{c|}{3.9}  & \multicolumn{1}{c|}{15.8}  & \multicolumn{1}{c|}{36.9}  & 51.4                        \\ \hline
\multicolumn{10}{c}{Image Restoration Methods}                                                                                                                                                                                                                                                                                                                                  \\ \hline
\multicolumn{1}{c|}{\multirow{4}{*}{CenterNet}} & \multicolumn{1}{c|}{\multirow{4}{*}{N}}         & \multicolumn{1}{c|}{BM3D}                     & \multicolumn{1}{c|}{\multirow{4}{*}{-}}  & \multicolumn{1}{c|}{50.9}        & \multicolumn{1}{c|}{10.4} & \multicolumn{1}{c|}{0.8}  & \multicolumn{1}{c|}{6.8}   & \multicolumn{1}{c|}{27.9}  & 86.0                        \\ \cline{3-3} \cline{5-10} 
\multicolumn{1}{c|}{}                           & \multicolumn{1}{c|}{}                           & \multicolumn{1}{c|}{Cycle-ISP~\cite{Rest_CycleISP}}                & \multicolumn{1}{c|}{}                    & \multicolumn{1}{c|}{56.7}        & \multicolumn{1}{c|}{10.6} & \multicolumn{1}{c|}{1.3}  & \multicolumn{1}{c|}{7.1}   & \multicolumn{1}{c|}{28.4}  & 84.5                        \\ \cline{3-3} \cline{5-10} 
\multicolumn{1}{c|}{}                           & \multicolumn{1}{c|}{}                           & \multicolumn{1}{c|}{IMDN (AS)~\cite{Light_weight_SR_LIMN}}                & \multicolumn{1}{c|}{}                    & \multicolumn{1}{c|}{46.1}        & \multicolumn{1}{c|}{9.8}  & \multicolumn{1}{c|}{0.5}  & \multicolumn{1}{c|}{6.0}   & \multicolumn{1}{c|}{27.0}  & 84.6                        \\ \cline{3-3} \cline{5-10} 
\multicolumn{1}{c|}{}                           & \multicolumn{1}{c|}{}                           & \multicolumn{1}{c|}{DAGL~\cite{Rest_DAGL}}                     & \multicolumn{1}{c|}{}                    & \multicolumn{1}{c|}{57.1}        & \multicolumn{1}{c|}{11.4} & \multicolumn{1}{c|}{1.2}  & \multicolumn{1}{c|}{7.2}   & \multicolumn{1}{c|}{28.8}  & 83.8                        \\ \hline
\multicolumn{10}{c}{Super Resolution Methods + Image Restoration Methods}                                                                                                                                                                                                                                                                                                       \\ \hline
\multicolumn{1}{c|}{\multirow{2}{*}{CenterNet}} & \multicolumn{1}{c|}{\multirow{2}{*}{N}}         & \multicolumn{1}{c|}{IMDN+DAGL}                & \multicolumn{1}{c|}{\multirow{2}{*}{x2}} & \multicolumn{1}{c|}{60.2}        & \multicolumn{1}{c|}{15.4} & \multicolumn{1}{c|}{3.5}  & \multicolumn{1}{c|}{15.2}  & \multicolumn{1}{c|}{27.9}  & 49.7                        \\ \cline{3-3} \cline{5-10} 
\multicolumn{1}{c|}{}                           & \multicolumn{1}{c|}{}                           & \multicolumn{1}{c|}{SWIN-IR+DAGL}             & \multicolumn{1}{c|}{}                    & \multicolumn{1}{c|}{59.3}        & \multicolumn{1}{c|}{15.9} & \multicolumn{1}{c|}{3.2}  & \multicolumn{1}{c|}{15.8}  & \multicolumn{1}{c|}{28.7}  & 49.9                        \\ \hline
\multicolumn{10}{c}{Different Training Schemes}                                                                                                                                                                                                                                                                                                                                 \\ \hline
\multicolumn{1}{c|}{\multirow{3}{*}{CenterNet}} & \multicolumn{1}{c|}{N}                          & \multicolumn{1}{c|}{\multirow{4}{*}{-}}       & \multicolumn{1}{c|}{\multirow{4}{*}{-}}  & \multicolumn{1}{c|}{42.2}        & \multicolumn{1}{c|}{14.5} & \multicolumn{1}{c|}{1.2}  & \multicolumn{1}{c|}{10.4}  & \multicolumn{1}{c|}{38.6}  & 87.3                        \\ \cline{2-2} \cline{5-10} 
\multicolumn{1}{c|}{}                           & \multicolumn{1}{c|}{L}                          & \multicolumn{1}{c|}{}                         & \multicolumn{1}{c|}{}                    & \multicolumn{1}{c|}{76.0}        & \multicolumn{1}{c|}{16.4} & \multicolumn{1}{c|}{1.5}  & \multicolumn{1}{c|}{12.4}  & \multicolumn{1}{c|}{41.3}  & \textbf{87.4}                        \\ \cline{2-2} \cline{5-10} 
\multicolumn{1}{c|}{}                           & \multicolumn{1}{c|}{\multirow{2}{*}{N+L}}       & \multicolumn{1}{c|}{}                         & \multicolumn{1}{c|}{}                    & \multicolumn{1}{c|}{76.6}        & \multicolumn{1}{c|}{16.5} & \multicolumn{1}{c|}{1.8}  & \multicolumn{1}{c|}{12.2}  & \multicolumn{1}{c|}{42.2}  & 87.3                        \\ \cline{1-1} \cline{5-10} 
\multicolumn{1}{c|}{$D_r$ + CenterNet}            & \multicolumn{1}{c|}{}                           & \multicolumn{1}{c|}{}                         & \multicolumn{1}{c|}{}                    & \multicolumn{1}{c|}{80.0}        & \multicolumn{1}{c|}{17.7} & \multicolumn{1}{c|}{\textbf{4.8}}  & \multicolumn{1}{c|}{15.8}  & \multicolumn{1}{c|}{41.0}  & 43.6                        \\ \hline
\multicolumn{10}{c}{Proposed Method}                                                                                                                                                                                                                                                                                                                                            \\ \hline
\multicolumn{1}{c|}{\textbf{RestoreDet} (w/o $D_t$)}       & \multicolumn{1}{c|}{\multirow{2}{*}{N+L}}       & \multicolumn{1}{c|}{\multirow{2}{*}{-}}       & \multicolumn{1}{c|}{\multirow{2}{*}{-}}  & \multicolumn{1}{c|}{79.8}        & \multicolumn{1}{l|}{17.9} & \multicolumn{1}{c|}{2.5}  & \multicolumn{1}{l|}{15.9}  & \multicolumn{1}{l|}{42.5}  & \textbf{87.4}                        \\ \cline{1-1} \cline{5-10} 
\multicolumn{1}{c|}{\textbf{RestoreDet} (w $D_t$)}         & \multicolumn{1}{c|}{}                           & \multicolumn{1}{c|}{}                         & \multicolumn{1}{c|}{}                    & \multicolumn{1}{c|}{\textbf{80.5}}        & \multicolumn{1}{l|}{\textbf{18.2}} & \multicolumn{1}{c|}{3.0}  & \multicolumn{1}{l|}{\textbf{16.4}}  & \multicolumn{1}{l|}{\textbf{43.0}}  & \textbf{87.4}                        \\ \hline
\bottomrule[1.2pt]
\end{tabular}
\end{adjustbox}
\caption{The detection result on MS COCO~\cite{coco_dataset} and KITTI~\cite{kitti_dataset} with randomly degradation (down-sample rate, kernel type and noise level) process. It can be seen that our RetoreDet get superior performance and decent inference speed among various methods. Best result are \textbf{high lighted}.}\label{tab:result_random}
\end{table*}

\section{Experiments and Details}

\begin{table*}[]
\centering
\begin{adjustbox}{max width= 1 \linewidth}
\begin{tabular}{l|c|c|cc|cccc|c}
\toprule[1.2pt]
\hline
\multirow{2}{*}{test set} & \multirow{2}{*}{model}              & \multirow{2}{*}{train set} & \multicolumn{2}{c|}{Pre-process to test set}                        & \multicolumn{4}{c|}{COCO}                                                                                    & \multirow{2}{*}{FPS (COCO)} \\ \cline{4-9}
                          &                                     &                            & \multicolumn{1}{c|}{method}                   & up-resolution       & \multicolumn{1}{c|}{AP}   & \multicolumn{1}{c|}{APs} & \multicolumn{1}{c|}{APm}  & APl                       &                             \\ \hline
\multirow{12}{*}{down 2}  & \multirow{9}{*}{CenterNet}          & \multirow{7}{*}{N}         & \multicolumn{1}{c|}{-}                        & -                   & \multicolumn{1}{l|}{17.7} & \multicolumn{1}{l|}{1.3} & \multicolumn{1}{l|}{13.6} & \multicolumn{1}{l|}{45.1} & 106.2                       \\ \cline{4-10} 
                          &                                     &                            & \multicolumn{1}{c|}{bicubic}                  & \multirow{4}{*}{x2} & \multicolumn{1}{c|}{19.5} & \multicolumn{1}{c|}{5.3} & \multicolumn{1}{c|}{20.1} & 33.1                      & 57.2                        \\ \cline{4-4} \cline{6-10} 
                          &                                     &                            & \multicolumn{1}{c|}{IMDN~\cite{Light_weight_SR_LIMN}}                     &                     & \multicolumn{1}{c|}{19.7} & \multicolumn{1}{c|}{5.4} & \multicolumn{1}{c|}{20.6} & 35.1                      & 56.8                        \\ \cline{4-4} \cline{6-10} 
                          &                                     &                            & \multicolumn{1}{c|}{HAN~\cite{SR_HAN}}                      &                     & \multicolumn{1}{c|}{19.2} & \multicolumn{1}{c|}{4.8} & \multicolumn{1}{c|}{19.1} & 32.9                      & 57.0                        \\ \cline{4-4} \cline{6-10} 
                          &                                     &                            & \multicolumn{1}{c|}{BSRGAN~\cite{SR_cvpr2021_random}}                   &                     & \multicolumn{1}{c|}{20.4} & \multicolumn{1}{c|}{5.8} & \multicolumn{1}{c|}{20.8} & 37.8                      & 56.5                        \\ \cline{4-10} 
                          &                                     &                            & \multicolumn{1}{c|}{BM3D}                     & \multirow{7}{*}{-}  & \multicolumn{1}{c|}{18.0} & \multicolumn{1}{c|}{1.2} & \multicolumn{1}{c|}{14.3} & 42.8                      & 105.6                       \\ \cline{4-4} \cline{6-10} 
                          &                                     &                            & \multicolumn{1}{c|}{DAGL~\cite{Rest_DAGL}}                     &                     & \multicolumn{1}{c|}{18.2} & \multicolumn{1}{c|}{1.6} & \multicolumn{1}{c|}{14.6} & 43.3                      & 106.3                       \\ \cline{3-4} \cline{6-10} 
                          &                                     & L                          & \multicolumn{1}{c|}{\multirow{5}{*}{-}}       &                     & \multicolumn{1}{c|}{19.1} & \multicolumn{1}{c|}{1.5} & \multicolumn{1}{c|}{15.3} & 46.1                      & 109.6                       \\ \cline{3-3} \cline{6-10} 
                          &                                     & \multirow{4}{*}{N+L}       & \multicolumn{1}{c|}{}                         &                     & \multicolumn{1}{c|}{19.1} & \multicolumn{1}{c|}{1.6} & \multicolumn{1}{c|}{15.5} & 46.0                      & 109.8                       \\ \cline{2-2} \cline{6-10} 
                          & $D_r$ + CenterNet                     &                            & \multicolumn{1}{c|}{}                         &                     & \multicolumn{1}{c|}{21.1} & \multicolumn{1}{c|}{\textbf{5.6}} & \multicolumn{1}{c|}{\textbf{21.1}} & 36.9                      & 39.1                        \\ \cline{2-2} \cline{6-10} 
                          & RestoreDet (w/o $D_t$)                &                            & \multicolumn{1}{c|}{}                         &                     & \multicolumn{1}{c|}{20.8} & \multicolumn{1}{c|}{2.8} & \multicolumn{1}{c|}{19.4} & 47.1                      & 108.8                       \\ \cline{2-2} \cline{6-10} 
                          & RestoreDet (w $D_t$)                  &                            & \multicolumn{1}{c|}{}                         &                     & \multicolumn{1}{c|}{\textbf{21.5}} & \multicolumn{1}{c|}{3.2} & \multicolumn{1}{c|}{20.8} & \textbf{47.3}                      & 108.7                       \\ \hline
\multirow{18}{*}{down 4}  & \multirow{14}{*}{CenterNet}         & \multirow{12}{*}{N}        & \multicolumn{1}{c|}{-}                        & -                   & \multicolumn{1}{c|}{7.3}  & \multicolumn{1}{c|}{0.0} & \multicolumn{1}{c|}{2.8}  & 30.1                      & 130.1                       \\ \cline{4-10} 
                          &                                     &                            & \multicolumn{1}{c|}{\multirow{2}{*}{bicubic}} & x2                  & \multicolumn{1}{c|}{12.9} & \multicolumn{1}{c|}{0.8} & \multicolumn{1}{c|}{8.7}  & 34.4                      & 107.2                       \\ \cline{5-10} 
                          &                                     &                            & \multicolumn{1}{c|}{}                         & x4                  & \multicolumn{1}{c|}{9.6}  & \multicolumn{1}{c|}{2.0} & \multicolumn{1}{c|}{9.5}  & 17.6                      & 57.2                        \\ \cline{4-10} 
                          &                                     &                            & \multicolumn{1}{c|}{\multirow{2}{*}{IMDN~\cite{Light_weight_SR_LIMN}}}    & x2                  & \multicolumn{1}{c|}{12.7} & \multicolumn{1}{c|}{0.7} & \multicolumn{1}{c|}{8.5}  & 35.1                      & 109.2                       \\ \cline{5-10} 
                          &                                     &                            & \multicolumn{1}{c|}{}                         & \multirow{3}{*}{x4} & \multicolumn{1}{c|}{10.3} & \multicolumn{1}{c|}{1.8} & \multicolumn{1}{c|}{10.5} & 16.9                      & 56.1                        \\ \cline{4-4} \cline{6-10} 
                          &                                     &                            & \multicolumn{1}{c|}{Real-SR~\cite{SR_real_images_1}}                  &                     & \multicolumn{1}{c|}{11.3} & \multicolumn{1}{c|}{2.0} & \multicolumn{1}{c|}{9.4}  & 17.4                      & 55.4                        \\ \cline{4-4} \cline{6-10} 
                          &                                     &                            & \multicolumn{1}{c|}{RRDB~\cite{RRDB}}                     &                     & \multicolumn{1}{c|}{8.9}  & \multicolumn{1}{c|}{1.9} & \multicolumn{1}{c|}{8.4}  & 16.2                      & 53.2                        \\ \cline{4-10} 
                          &                                     &                            & \multicolumn{1}{c|}{\multirow{2}{*}{BSRGAN~\cite{SR_cvpr2021_random}}}  & x2                  & \multicolumn{1}{c|}{13.3} & \multicolumn{1}{c|}{1.3} & \multicolumn{1}{c|}{11.3} & 38.9                      & 104.2                       \\ \cline{5-10} 
                          &                                     &                            & \multicolumn{1}{c|}{}                         & x4                  & \multicolumn{1}{c|}{11.2} & \multicolumn{1}{c|}{\textbf{2.1}} & \multicolumn{1}{c|}{10.9} & 19.8                      & 58.1                        \\ \cline{4-10} 
                          &                                     &                            & \multicolumn{1}{c|}{BM3D}                     & \multirow{8}{*}{-}  & \multicolumn{1}{c|}{7.6}  & \multicolumn{1}{c|}{0.0} & \multicolumn{1}{c|}{3.2}  & 29.8                      & 119.5                       \\ \cline{4-4} \cline{6-10} 
                          &                                     &                            & \multicolumn{1}{c|}{Cycle-ISP~\cite{Rest_CycleISP}}                &                     & \multicolumn{1}{c|}{8.0}  & \multicolumn{1}{c|}{0.1} & \multicolumn{1}{c|}{3.5}  & 32.5                      & 119.7                       \\ \cline{4-4} \cline{6-10} 
                          &                                     &                            & \multicolumn{1}{c|}{DAGL~\cite{Rest_DAGL}}                     &                     & \multicolumn{1}{c|}{8.2}  & \multicolumn{1}{c|}{0.1} & \multicolumn{1}{c|}{3.3}  & 33.2                      & 120.1                       \\ \cline{3-4} \cline{6-10} 
                          &                                     & L                          & \multicolumn{1}{c|}{\multirow{5}{*}{-}}       &                     & \multicolumn{1}{c|}{10.5} & \multicolumn{1}{c|}{0.1} & \multicolumn{1}{c|}{5.0}  & 38.3                      & 123.3                       \\ \cline{3-3} \cline{6-10} 
                          &                                     & \multirow{5}{*}{N+L}       & \multicolumn{1}{c|}{}                         &                     & \multicolumn{1}{c|}{10.7} & \multicolumn{1}{c|}{0.1} & \multicolumn{1}{c|}{5.2}  & 38.7                      & 122.1                       \\ \cline{2-2} \cline{6-10} 
                          & $D_r$ + CenterNet                     &                            & \multicolumn{1}{c|}{}                         &                     & \multicolumn{1}{c|}{13.4} & \multicolumn{1}{c|}{1.6} & \multicolumn{1}{c|}{11.4} & 33.1                      & 45.2                        \\ \cline{2-2} \cline{6-10} 
                          & RestoreDet (w/o $D_t$)                &                            & \multicolumn{1}{c|}{}                         &                     & \multicolumn{1}{c|}{11.7} & \multicolumn{1}{c|}{0.4} & \multicolumn{1}{c|}{8.9}  & 41.0                      & 120.7                       \\ \cline{2-2} \cline{6-10} 
                          & \multirow{2}{*}{RestoreDet (w $D_t$)} &                            & \multicolumn{1}{c|}{}                         &                     & \multicolumn{1}{c|}{12.0} & \multicolumn{1}{c|}{0.6} & \multicolumn{1}{c|}{9.1}  & \textbf{41.8}                      & 120.7                       \\ \cline{4-10} 
                          &                                     &                            & \multicolumn{1}{c|}{bicubic}                  & x2                  & \multicolumn{1}{c|}{\textbf{14.3}} & \multicolumn{1}{c|}{1.5} & \multicolumn{1}{c|}{\textbf{12.6}} & 34.9                      & 106.0                       \\ \hline
\bottomrule[1.2pt]
\end{tabular}
\end{adjustbox}
\caption{Detection results on MS COCO~\cite{coco_dataset} with fixed downsampling rate (2/4). Our RestoreDet achieves superior performance among the same size evaluation compared to alternatives. Best result are \textbf{high lighted}.}\label{tab:result_fixed}
\end{table*}

\subsection{Datasets and Implementation Details}
\label{sec:train_details}

We build our framework based on the open-source object detection toolbox \textit{mmdetection} ~\cite{mmdetection}. Two well-known object detection datasets, KITTI ~\cite{kitti_dataset} and MS COCO ~\cite{coco_dataset}, have been used to evaluate our method. Throughout the experiments, the  backbone, ResNet-18 ~\cite{resnet}, is  pre-trained on ImageNet ~\cite{imagenet} dataset and the training settings adopt the same data augmentation methods (random crop, multi-size, random flip).  All input images are resized to $512 \times 512$ in the training stage. 


KITTI ~\cite{kitti_dataset} is a popular small object detection dataset for autonomous driving. For KITTI dataset, we evaluate $car$ class. All of the models are trained on KITTI train set and evaluated on KITTI validation set. They have been trained on a single RTX 6000 GPU for 110 epochs with SGD optimizer. The batch size is set to 16 and the momentum and weight decay are set to 0.9 and 1e-4. The initial learning rate of encoder, detection and data restoration decoder is 1e-3, while the transformation decoder part is 1e-4. The learning rate warms up ~\cite{SGD_warmup} at first 500 iterations and decays to one-tenth at 100 and 105 epochs. For evaluation, we report the mean average precision (mAP) rate at IoU threshold of 0.5.

MS COCO ~\cite{coco_dataset} is a popular large dataset, which contains over 10 million images with 80 categories. In the experiment, we evaluate all 80 classes. For COCO dataset training, all models are trained on COCO 2017 train set and evaluated on COCO 2017 validation set.  They have been trained on 4 RTX 6000 GPUs for 140 epochs, with batch size 32. The initial learning rate is the same as KITTI setting and warms up at the first 500 iterations, then decays to one-tenth at 90 and 120 epochs. For evaluation, we report evaluation metric of COCO indexes as: $\left\{ AP, AP_S, AP_M, AP_L \right\}$ which shows  object detection performance on different scales. 

In the testing stage, all the results are tested on a single RTX 6000 GPU. We compare the speed by reporting the  frames per second (FPS) in the experiments of COCO dataset.

\begin{figure*}
    \centering
    \includegraphics[width = 17.0cm, height = 13.0cm]{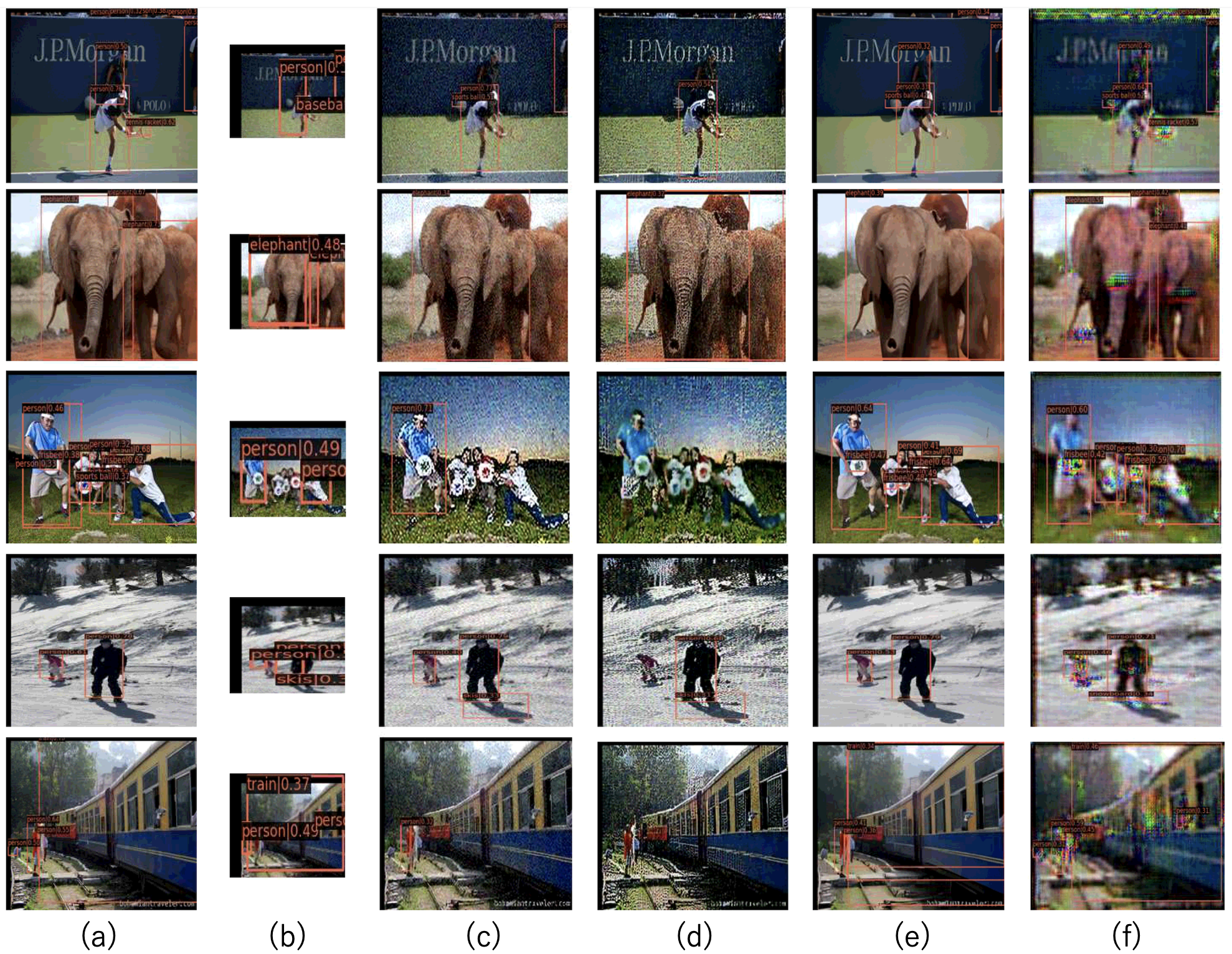}
    \caption{Exemplar detection results on MS COCO 2017 dataset~\cite{coco_dataset}. (a)/(b) is CenterNet trained on normal images and tested on normal/degraded down4 testset, (c)/(d)/(e) is CenterNet tested on the degraded image restored by individual SR algorithm RRDB~\cite{RRDB}/Real-SR~\cite{SR_real_images_1}/BSRGan~\cite{SR_cvpr2021_random}. (f) is the detection result of our RestoreDet and we use the output of ARRD $D_r$ as background images.}
    \label{fig:results}
\end{figure*}

\subsection{Experiment Results}


To synthesize different degrees of degraded images, we generate the LR image with Eq.~\ref{eq:degradation_model} with random down-sampling rate (random chosen from a uniform distribution $U(1, 4)$), random blur kernel (isotropic Gaussian kernel $k_{iso}$, anisotropic Gaussian kernel $k_{aniso}$ and no blur kernel $k_{none}$), random noise level (AWGN noise with variance randomly chosen from $U(0, 25/255)$).



At first we compare our methods with SOTA image restoration methods, as shown in Table.\ref{tab:result_random}. For SR methods, the tested images would be enhanced by interpolation method, SOTA non-blind SR models~\cite{SR_HAN, SR_PAN, Light_weight_SR_LIMN, SR_RCAN, SR_SWIN-IR}, SOTA blind SR models~\cite{SR_real_images_1, RRDB, SR_cvpr2021_random, Siggraph_blind_SR}. We also compare with latest image restoration methods (do not change image resolution) like~\cite{Rest_CycleISP, Rest_DAGL}. From Table~\ref{tab:result_random}, we find that enhanced higher resolution (either by interpolate or SR pre-process) improves the detection performance on small objects, but has a negative impact on the large object as well as the inference speed. Super-resolving the image to an excessively resolution will also lead to very undesirable results. Compared to other non-blind SISR methods~\cite{Light_weight_SR_LIMN,SR_PAN,SR_HAN,SR_SWIN-IR},  blind SR methods ~\cite{Siggraph_blind_SR, SR_cvpr2021_random} show better improve in object detection under diverse degradation. 


We also train the  CenterNet on different training schemes as shown in Table.\ref{tab:result_random}.  $N$ means the original normal training set. $L$ corresponds to the LR images from the random degradation operations in Sec.\ref{sec:degradation}.  $N+L$ denotes a  mixed of normal and LR images. To compare with the structure like~\cite{SR_object_detection}, we use ARRD $D_r$ as the pre-processing module for CenterNet  and jointly  optimized them. The performance is denoted as the $D_r +$ CenterNet in Table.\ref{tab:result_random}. Although  $D_r +$  performs well on small object detection, it  consumes much time in inference. For our RestoreDet's evaluation, we separately test RestoreDet trained with and without transformation decoder $D_t$ as (w/o $D_t$) and (w $D_t$) in Table.\ref{tab:result_random}.
After all, our RestoreDet (w $D_t$) consistently achieves the SOTA detection accuracy and highest FPS. As shown in in Fig.\ref{fig:results} (f), the ARRD's outputs (restored images $\hat{x}$) are more machine vision oriented and exhibit artifacts around the center of the object.


To better verify the effectiveness of our model, similar to~\cite{SR_object_detection}, we also evaluate our performance on COCO with a fixed down-sampling setting (2 and 4). We need to emphasis that the RestoreDet model here is trained under previous setting. Since the  real-world down-sampling condition is unknown, our RestoreDet should be able to automatically adapt to a new setting. As shown in Table~\ref{tab:result_fixed}, our method continuously gets the best result in the same scale evaluation and different training schemes. Our method also keeps a fast computational speed because we avoid reconstructing the HR details.

\subsection{Ablation Study}
\begin{table}[!htb]
\centering
\begin{adjustbox}{max width= 1.5 \linewidth}
\begin{tabular}{c|c|c}
\hline
\hline
(resolution: 1$\sim$4) & AP   & Recall \\ \hline
CenterNet (N)          & 42.2 & 58.8   \\ \hline
CenterNet (L)          & 76.0 & 85.7   \\ \hline
+ $D_t$ (id = 1)           & 77.2 & 87.7   \\ \hline
+ $D_t$ (id = 2)           & 77.7 & 88.5   \\ \hline
+ $D_t$ (id = 3)         & 76.8 & 87.0   \\ \hline
+ $D_r$                  & 79.8 & 90.0   \\ \hline
+ $D_r$ + $D_t$ (id = 2)   & \textbf{80.5} & \textbf{92.2}   \\ \hline
\hline
\end{tabular}
\end{adjustbox}
\caption{Ablation study on KITTI dataset~\cite{kitti_dataset}}
\label{tab:Ablition_study}
\end{table}

To evaluate each component in our RestoreDet, we make an ablation study on KITTI dataset~\cite{kitti_dataset}. The evaluation metric is AP rate and Recall rate while the training setting is the same as the above section. We report the evaluation results  in Table.~\ref{tab:Ablition_study}.

We first trained the vanilla CenterNet on the original/degraded KITTI dataset and evaluated it on the degraded (down $1 \sim 4$) KITTI dataset as CenterNet (N) / CenterNet (L) in Table.~\ref{tab:Ablition_study}. The augmented degraded training data has greatly improved the detection results. Then we  added the transformation decoder $D_t$ and ARRD $D_r$. For transformation decoder $D_t$, we evaluate the variants when $D_t$  is connected to different feature stages in ResNet backbone ( $id = 1, 2, 3$ corresponds to the ResNet $2, 3, 4$ stage output, respectively). $+D_r$ refers to  adding ARRD and different level feature combination process in the network. The final $+ D_r + D_t$ is our full RestoreDet architecture, which achieves the best performance.


\section{Conclusion}
In this paper, we propose a novel self-supervised framework, RestoreDet, to handle object detection for degraded low resolution images. To capture the dynamics of feature representations under diverse resolution and degradation conditions, we propose a degradation equivariant representation that is generic and could be implemented on popular detection architectures. To further combine the strength of the existing progress on super resolution (SR), we  also introduce an arbitrary-resolution restoration decoder that supervises the latent representation to preserve the visual structure. The extensive experiments demonstrate that  our RestoreDet achieves SOTA results on  two mainstream public datasets among different degradation conditions (resolution, noise and blur).


{\small
\bibliographystyle{ieee_fullname}
\bibliography{egbib}
}

\end{document}